\documentclass[conference]{IEEEtran}
\usepackage{cite}
\usepackage{hyperref}
\hypersetup{
    colorlinks=false,
    linkcolor=blue,
    filecolor=magenta,      
    urlcolor=blue,
    pdftitle={},
    pdfpagemode=FullScreen,
    }
\usepackage{amsmath,amssymb,amsfonts}
\usepackage[utf8]{inputenc}

\usepackage{algorithmic}
\usepackage{graphicx}
\usepackage{textcomp}
\usepackage{xcolor}
\usepackage{acronym}
\usepackage{tabularx}
\usepackage{tikz}
\usetikzlibrary{calc}

\def\BibTeX{{\rm B\kern-.05em{\sc i\kern-.025em b}\kern-.08em
    T\kern-.1667em\lower.7ex\hbox{E}\kern-.125emX}}

\usepackage{array,booktabs}
\newcommand {\otoprule}{\midrule [\heavyrulewidth]}
\newcolumntype {+}{ >{\global\let\currentrowstyle\relax}}
\newcolumntype {^}{ >{\currentrowstyle }}
  \newcommand {\rowstyle}[1]{\gdef\currentrowstyle{#1} %
  #1\ignorespaces
  }
\newcommand{\tabhead}{\rowstyle{\bfseries}}

\begin{document}

\title{Towards a trustworthy, secure and reliable enclave for machine learning in a hospital setting:\\  {The \acf{EMCP}}}

\author{
    \IEEEauthorblockN{%
    Hendrik F.R. Schmidt\IEEEauthorrefmark{1}, 
    J\"{o}rg Schl\"{o}tterer\IEEEauthorrefmark{1}\IEEEauthorrefmark{2}\IEEEauthorrefmark{3},
    Marcel Bargull\IEEEauthorrefmark{1}, 
    Enrico Nasca\IEEEauthorrefmark{1}\IEEEauthorrefmark{3}, 
    {Ryan Aydelott\IEEEauthorrefmark{1}},}
    Christin Seifert\IEEEauthorrefmark{1}\IEEEauthorrefmark{2}\IEEEauthorrefmark{3}\IEEEauthorrefmark{4},
    Folker Meyer\IEEEauthorrefmark{1}\IEEEauthorrefmark{2}\IEEEauthorrefmark{4}
    \IEEEauthorblockA{\IEEEauthorrefmark{1}Institute for Artificial Intelligence in Medicine, University Hospital Essen, Essen, Germany
    \\\{firstname.lastname\}@uk-essen.de}
    \IEEEauthorblockA{\IEEEauthorrefmark{2}University of Duisburg-Essen, Essen, Germany
    \\\{firstname.lastname\}@uni-due.de}
        \IEEEauthorblockA{\IEEEauthorrefmark{3}Cancer Research Center Cologne Essen (CCCE), Essen, Germany}
    \IEEEauthorblockA{\IEEEauthorrefmark{4}Co-senior authors}
}

\maketitle

\begin{tikzpicture}[remember picture, overlay]
\node at ($(current page.north) + (0,-0.5in)$) {\footnotesize Author Preprint, \textcopyright 2021 IEEE, Proc. IEEE CogMI 2021};
\end{tikzpicture}

\acresetall
\begin{abstract}
AI/Computing at scale is a difficult problem, especially in a health care setting. We outline the requirements, planning and implementation choices as well as the guiding principles that led to the implementation of our secure research computing enclave, the \ac{EMCP}, affiliated with a major German hospital. Compliance, data privacy and usability were the immutable requirements of the system. We will discuss the features of our computing enclave and we will provide our recipe for groups wishing to adopt a similar setup.\footnote{\label{fn:ansible-scripts}The Ansible project is available from \url{https://github.com/IKIM-Essen/EMCP-config}}
\end{abstract}


\section{Introduction}
The use of scientific computing has great potential for health care; in fact currently researchers are exploring computing approaches to classify potential melanoma~\cite{DermAssist}, detect breast cancer~\cite{mammo} and personalize treatment~\cite{RLinHealthSurvey}. \ac{ML} approaches are opening new venues for applied clinical research, while the advent of tooling from computer science is triggering a digital transformation of health care and health care-related research. 
Colocation of computational research labs and health care IT have fundamentally different technical requirements. This goes beyond traditional IT resources in health care and moreover does not allow for separation of concerns between clinical services and research. Health care IT with a myriad of different systems and many users of different skill sets cannot easily accommodate a computational research platform as they have very specific requirements along with vastly different user expectations. 

Efficient computational research requires rapid evaluation of published software tools and libraries, as well as scripting on top of a heterogeneous, often changing set of dependencies.
In contrast, clinical environments are not typically supportive of individual researchers installing software without the need for documentation and/or management or IT staff signoff.
This is the result of of the fact that health care data (particularly patient data) is protected by a number of laws and regulations in Germany. To comply with these laws, a significant number of technical measures must be taken to insure that access is controlled and data on the system is protected. It is not an understatement to declare that hospitals take IT security and the rule of parsimonious access quite seriously.

While hospitals in most developed countries are IT rich environments with an abundance of systems, it is useful to point out that those IT infrastructures were not designed to facilitate research computing, let alone resource-intensive computing. More often than not, the research environment is a researcher's desktop system, which typically has the least amount of restrictions and access controls and is rarely the subject of a professional audit. A department conducting large scale \acf{ML} or \acf{DL} experiments or resource-intensive computing clearly is not a good fit for a typical hospital. Even the largest research hospital can struggle to comply with functional requirements, access controls and audit procedures for a system focused on research, as they differ vastly both in purpose and structure from the systems that traditional IT health care playbooks were written for.

Research computing often requires access to resources significant in both quantity and novelty. This includes \acp{GPU}, \acp{FPGA} and \acp{ASIC} (including custom silicon accelerator hardware), as well as diverse computational libraries and frameworks with specific versioning requirements. Altogether this can pose a formidable challenge to even the most advanced health care IT environments when constrained by operational procedures written againsts non-research based systems.

We therefore conclude that a separate space is required to conduct large scale computational research. This environment is specifically crafted to facilitate research productivity. Since its user base and function is vastly different, many of the traditional operational constraints imposed by hospital security or operational requirements can be avoided. One of the largest hurdles to this type of system is the principle of parsimonious access to data or systems access privileges. This hampers researchers as installing software libraries required for research is rightfully considered undesirable for highly regulated and access-controlled hospital environments.
Consequently, we seek an environment that removes those barriers by enabling researchers to install software and frameworks themselves, while simultaneously maintaining a reasonable level of overall security in the infrastructure.
\emph{In summary, a dedicated \ac{ML} enclave is required adjacent to the hospital IT systems.}

\section{Example Use Cases}
In this section we exemplify some computing use cases and outline their requirements on a computing infrastructure.

\subsection{Understanding Hospital Patient Flows} 
One key indicator for quality of health care is the quality of patient journeys. A patient's journey through the hospital might deviate from the guidelines for multiple reasons, such as unforeseeable events and the complexity of the organization. \acp{EHR} contain information about events related to diagnostics and treatment of patients and can serve as input for retracing a patient's journey~\cite{Marazza2020_Patient-Process-Models}. The application of tools to compute a patient's journey from observational data is called process mining. Process mining tools require event types (e.g., ``an MRI was performed") and an associated time stamp. Such an application does not require personal data, but the data integrated from multiple hospital departments and their subsystems (e.g., pathology and radiology departments), faciliate the need for specific computing environments to install and run these process mining tools in addition to providing a data anonymization toolchain.

\subsection{Data-driven Treatment Effect Prediction in Oncology}
Oncological research on treatment effects relies on a structured, summarized patient history (e.g.,~\cite{Wiesweg2021}), that contains treatment events from the first diagnosis until patient discharge. Currently, these structured reports are created by clinicians by manually coalescing information from multiple IT systems (this may contain, but shall not be limited to: computertomography reports, chemotherapy information, treatment plans, other clinical reports) and resolving contradictions from multiple sources. 
While the current process builds on abstractions on multiple levels, a purely data-driven approach would need to integrate raw data from various sources, ranging from images (magnetic resonance imaging, ultrasound, digital histopathology, ...) over structured data such as lab results, to natural language contained in reports.
The challenges posed by these data sources comprise variety, volume and computational effort.
In particular whole slide images, as used in digital histopathology require vast amounts of storage capacity, whereas the training time of state of the art natural language processing models is well within thousands of \ac{GPU} hours.
These traditional "Big Data" problems to derive both context and value from diverse sources are significantly hampered by the need for patient privacy, which requires a much more crafted approach.

\subsection{Using Microbiome Data to Inform the Selection of Antibiotics to Treat Infection}
With metagenomic sequencing (see e.g.,~\cite{thomas2012}) characterizing a patients microbiome and thus any potential antibiotics resistance genes is well within the capabilities of modern science. However, characterizing the myriad of bacterial genomic fragements that metagenomic sequencing produces requires significant computational resources. While workflows exist already (e.g.,~\cite{meyer2017}) their execution requires either containerized execution environments at scale or the ability to install bioinformatics software (e.g, via Bioconda~\cite{bioconda}) neither of these are a good fit for the highly regulated hospital environment. In addition, few of these environments can provide access to enough systems (scale) needed to process current microbiome data in a timely manner. Due to the relative novelity and the lack of mature software systems, outsourcing the computational work can create its own challenges as it would create yet another black box introducing yet another set of obstacles for scientific discovery. Performing metagenome analysis to analyze microbiome data will require about 400 CPU core hours per gigabyte~\cite{meyer2017} of data assuming the least resource intensive workflows. Additionally this requires tens of thousands of files to be stored, many of them temporarily, all while requiring the installation of several dozen bioinformatics analysis tools, many of which have significant dependencies in the form of libraries etc.

 \section{The Archetypical User} 
 Scientific computing users are comprised of a rather diverse group of (PhD) students, software developers, computer scientists, clinicial scientists, bio-medical researchers and various (advanced) users.
 In an attempt to introduce the prototypical user we identified diverging requirements as the most common characteristic -- not only across user groups, but also for the same single user.
 For example, when a machine learning researchers start prototyping, they want to run state of the art \ac{ML} software and typically either are not proficient, nor should they spend their time dealing with `low-level' setup details (e.g., \ac{GPU} drivers, \ac{CUDA} versions) that are traditionally the realm of system architects/integrators. In light of this, they should be provided with a ready-to-use environment (e.g., as provided by Google Colab~\cite{colab} -- just enter PyTorch, Tensorflow, or code of their favorite framework in a Jupyter Notebook and hit run). 
 As the \ac{ML} prototype models become more advanced, they may want to optimize model distribution and communication over multiple computation nodes, requiring access to low-level routines.
 In addition, our typical ML user has several dozen terabytes of image data, as well as text and structured data that needs to be extracted from a myriad of hospital business systems. 
 We discovered that there is \emph{not a single archetypical user} but rather a whole \emph{host of usage patterns} that we need to support. 
 Consequently, our user model includes a number of perspectives and also a number of services required to meet our researchers' needs.
 
 \section{Regulatory constraints}
Data access in a hospital environment in Germany is regulated by a number of laws and regulations. IT security in larger hospitals in Germany is regulated by KRITIS (\S 8a IT-SiG, \S 6 BSI-KritisV), the German social code (\S 75b SGB V, \S 75c SGB V), national data privacy law (\S 22 section 2. BDSG), and finally the \ac{GDPR}~\cite{EUdataregulations2018} or its German implementation. There are also additional codes covering the data protection requirements that vary by federal state (LDSG and LKHG).

While it is not approrpriate to summarize the legal aspects here, the basic take home is parsimonious access, requirements for authentication and a need for documentation of procedures and access is an important part of any system. The authors deem data privacy to be a very important component of civil rights. However, it can be argued that compliance with a complex set of laws (see above) has not led to a hospital IT landscape conducive to a research-first approach. Certain components of the governing rules lead to unforseen side-effects now that computational science is trying to get a foothold in the hospital system: e.g., the need to obtain an ethics vote prior to any use of data with \ac{PHI} ultimately leads to a de-facto ban on explorative studies.

Working with our organizational data privacy officer we devised a plan to create an environment that allows the application of modern computing tools to (bio-)medical data in a hospital environment.
By removing the \ac{PHI} from the data sets, data is considered no longer protected by the above mentioned laws and regulations. We therefore have reduced the regulatory burden and unlocked the full repertoire of scientific computing.

While we have removed \ac{PHI} from the data, we nevertheless acknowledge the potential for misuse of the data and therefore implement robust IT security policies for the enclave.

 \section{System Requirements}
 The system needs to provide state of the art IT security and minimize compliance constraints as much as possible.
 Any future solution needs to comply with the various {legal requirements} (in our case EU GDPR, German federal \S 22 Abs 2 BDSG and state LDSG)\footnote{For an international audience: The rule set is similar to the HIPAA regulatory framework in the U.S.}. 
 However, in addition to complying with the legal framework, we need to ensure that the solution is deemed {socially acceptable} by both our peers inside the hospital as well as the general public, thus processing of \ac{PHI} needs to be limited to an absolute minimum or avoided altogether.
 Social acceptance has another facet: {developers} need to find the working environment acceptable to achieve high performance. Any solution must make available modern computational abstractions that render developers and scientists productive as well as limit the amount of oversight required, e.g., for PhD students when working with data. Modern computing abstractions (e.g., container registries, Helm Charts, existing \ac{ML}/\ac{DL} systems) are critical aspects of performance in an organisation that both develops and maintains software environments.
 The system needs to provide {data ingress and egress} solutions.
 And finally, ideally the system itself will {utilize the expertise} of our scientists and developers and allow us to evolve the system over time.
 
As our users' requirements change over time, we need to devise a platform for service provisioning rather than a fixed set of services. The list of services is certain to grow over time, but initially we included the following services in our plans:
\begin{itemize}
    \item   \textbf{File services}, i.e., secure \ac{NFS} services inside the enclave and \ac{SMB} services to specific medical research devices outside patient care.
    \item  \textbf{Object store}, i.e., secure S3 services inside and outside the enclave.
    \item \textbf{Scheduling and resource management} for various classes of computational resources, e.g., specific \ac{GPU} families.
    \item \textbf{Web-based user interfaces}, such as Jupyter~\cite{jupyter}, an interactive data science platform, requiring little or no command-line experience.
    \item \textbf{Package management} for scientific software. A myriad of scientific software systems exist as pre-packaged Conda packages.
    \item \textbf{Version control software}, such as Git in the form of CI/CD enabled tools such as GitLab~\cite{gitlab}.
    \item \textbf{Team collaboration software}, such as Rocket.Chat~\cite{rocketchat} or Mattermost~\cite{mattermost}. 
\end{itemize}

\section{System Design} 
\subsection{Solution Sketch}

\begin{figure*}[tbp]
    \centering
    \includegraphics[width=0.7\textwidth]{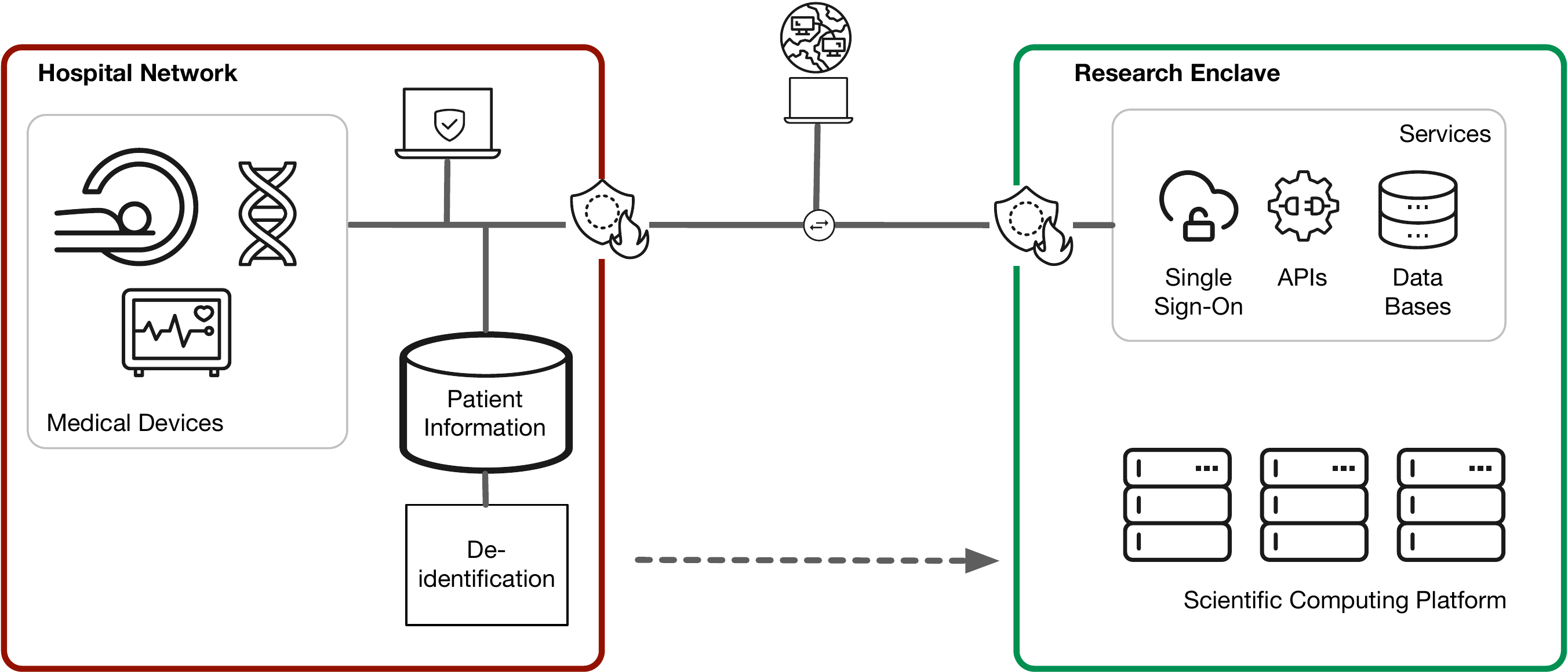}
    \caption{The research computing enclave (right) is adjacent to the hospital network (left) but separated from it via a firewall. Personal information is de-identified before transfer to the enclave.}
    \label{fig:IKIM_network}
\end{figure*}

We use Linux/UNIX as a scientific computing platform.
As international \textit{cloud} solutions, providing \ac{IaaS}~\cite{NistCloud2011} are deemed not acceptable for patient data both socially and legally in Germany, an on-premise solution was chosen. 

The systems are deployed in a dedicated network segment separate from the hospital and the internet (cf. \autoref{fig:IKIM_network}), with incoming access possible via SSH into a bastion host with internet connectivity. 
Multi-factor authentication is required for access to the enclave. 
A proxy service enables internet access via HTTP inside the enclave for system updates and maintenance.

To reduce the compliance load as well as the potential for disaster should anything go wrong, we opted to de-identify data prior to storage inside the enclave. 
De-identified data will be copied into the enclave on demand to enable scientific computing on said data. 
In addition to data from patient care, anonymous research data is also abundant in our hospital environment and we will integrate equipment across campus into the research enclave by establishing dedicated \acp{VLAN} connected to our file service to enable the flow of research data into the enclave.

While container-based approaches for research are gaining momentum, \textit{traditional} cluster computing approaches with shared file systems and centralized identity management are needed to assist at least some researchers.
Therefore, we decided to support both by providing shell access to a managed cluster, including rootless containers (to prevent priviledge escalation) and providing access to a bare metal provisioning service.

Finally, since we anticipate moving from our on-premise environment to a third-party \ac{IaaS} platform, we use an \ac{IaC} stance, coding and documenting the environment in a single source code repository using Ansible~\cite{ansible}. 
The use of a popular tool like Ansible enables us to share the workload of establishing the enclave between several developers with and without operational roles in running said enclave.

\subsection{Architecture Details}
In the following, we describe and motivate our implementation choices as well as putting them in context.

\subsubsection*{Physical / Network Level Separation}
Network level separation is almost a requirement to avoid the significant restrictions imposed by the German legal and compliance framework. A well-defined logical location outside the hospital proper enables more flexibility and reduces the required compliance workload. 

\subsubsection*{System Security, Firewall, \ac{2FA}}
While data in the enclave is assumed to be free of \ac{PHI} we nevertheless acknowledge that even in de-identified form, the data still represents a significant value as well as a potential threat to confidentiality. 
As a consequence, we chose to set up the system in a separate network location, protect it via a dedicated firewall system and require \ac{2FA} for access to the entire system. Hardware-based \ac{2FA} is currently considered best practice \cite{Colnago} for securing valuable data and computing assets.

\subsubsection*{On-Premise and Bare Metal Provisioning Service}
While renting access to computing equipment and storage is certainly de rigueur currently (especially with smaller scale/experimental test beds), for a production system both the scale of computing resources required along with the vast amount of data that needs be stored/accessed quickly made the decision to go with a self-hosted, on-premise solution an easy one. The fact that our hospital had a fully equipped but under-utilized data center removed some of the expenses traditionally incurred for a large on-premise hosted solution.
We chose \ac{MAAS}~\cite{MAAS} as a bare metal provisioning system with an end-user interface as the basis of our enclave, this provides an environment similar to current popular \textit{cloud} services, giving users an immediate familiarity. \ac{MAAS} further improves performance against most cloud systems as it does not have the performance limitations that virtualization layers typically impose via abstraction layers, which can be a burden to scientific computing performance.

\subsubsection*{Infrastructure as Code}
While we recognize that currently the financial and legal situation clearly favors on-premise solutions, we nevertheless acknowledge that in the future our solution might shift and we therefore decided to use an \ac{IaC} approach allowing us to migrate the entire setup to future platforms.
Furthermore this approach increases resilience of the overall computing infrastructure in case of hardware failure as well as facilitates upscaling as required.
We decided to implement our infrastructure as a series of Ansible roles, available widely across the institute in a GitHub repository.\textsuperscript{\ref{fn:ansible-scripts}}
We invite pull requests by all, utilizing all available expertise rather than frustrating end users with restrictions.
Importantly, the infrastructure is documented and pull requests for changes are enabled at the same time.
Due to using version control, every change to the infrastructure is transparent and can be rolled back on demand.
Using \ac{MAAS} as the bare metal provisioning service and an \ac{IaC} approach, we were able to quickly establish basic functionality and work together on an environment conducive to high research productivity.

\subsubsection*{File services (\ac{NFS}, \ac{SMB}), object storage and single sign-on}
Easily accessible file services covering both the \textit{traditional} (\ac{NFS}, \ac{SMB}) and object storage flavors enable a number of computational abstractions and provide ease of use for data transfer and use. We decided to install a server to share storage for ongoing research via \ac{NFS}v4, Samba~\cite{SAMBA} and another server acting as both S3-compatible object storage in/egress service and archive using MinIO~\cite{minio}.
Local disk caching compensates for \ac{NFS} performance issues.
To fully utilize the data storage services and enhance the overall utility of the system we chose a FreeIPA~\cite{freeipa} server as our \ac{LDAP} single sign-on system to centrally manage users and access credentials.

\subsubsection*{A Platform for Container Orchestration -- Kubernetes}
Provisioning of services such as databases, central applications, monitoring and messaging buses etc. in a cost effective and resilient way is currently almost synonymous with using Kubernetes~\cite{K8S}. The Helm~\cite{HELM} package manager provides pre-configured production-grade deployments for many data, communication or \ac{ML}/\ac{DL} services. The Kubernetes system itself is implemented via the open source version of 
Rancher~\cite{rancher} and provides local, redundant on-system storage to efficiently implement large and small-scale databases that would otherwise incur significant latency and bandwith limitation if they were implemented via a network-attached storage.
Furthermore, Rancher enables us to easily deploy multiple separate Kubernetes clusters for different use cases, such as infrastructure services on one side and scientific applications on the other side. 
This enables us to limit access to the underlying infrastructure these services are deployed on, thus increasing the protection of critical central services such as monitoring.

\subsubsection*{Bringing Together Cluster and Container-Based Computing}
An analysis of end-user expections and requirements made it clear that both, container-based approaches and \textit{traditional} cluster computing approaches with shared file systems and centralized identity management are needed.

Therefore we devised two modes of access to the system: i) a Linux cluster accessible via a login server providing shell access including rootless containers and ii) access to the bare metal provisiong service enabling users to provision their own nodes with full root access. Importantly, users can choose to self-administer their own nodes, using the Ansible recipes provided for cluster nodes, losing only the access to the \ac{NFS} file service. 
Both in the case of i) and in the case of ii) limitations are necessary to prevent priviledge escalation. For mode i) it must be ensured, that users inside a container are not able to alter or write files to the host with arbitrary UIDs\footnote{\url{docs.docker.com/engine/security/userns-remap/}}. 
Conversely, users in mode ii) must not be able to access the NFS file service, as they may assume the role of any user through local root privileges.\footnote{The authors are aware that using Kerberized \ac{NFS} would solve this problem but also increase complexity. The available S3-compatible object storage is the primary storage for users with independent platforms.} This way it would be possible to bypass access control and inflict arbitrary damage on the overall system.

\subsection{Planning for Data Ingress and Egress}
Access to patient and research data is essential for scientific computing. Hence, building data conduits that work for the intended audience and yet guarantee privacy is critical.

\begin{figure}[tbp]
    \centering
\includegraphics[width=0.9\columnwidth]{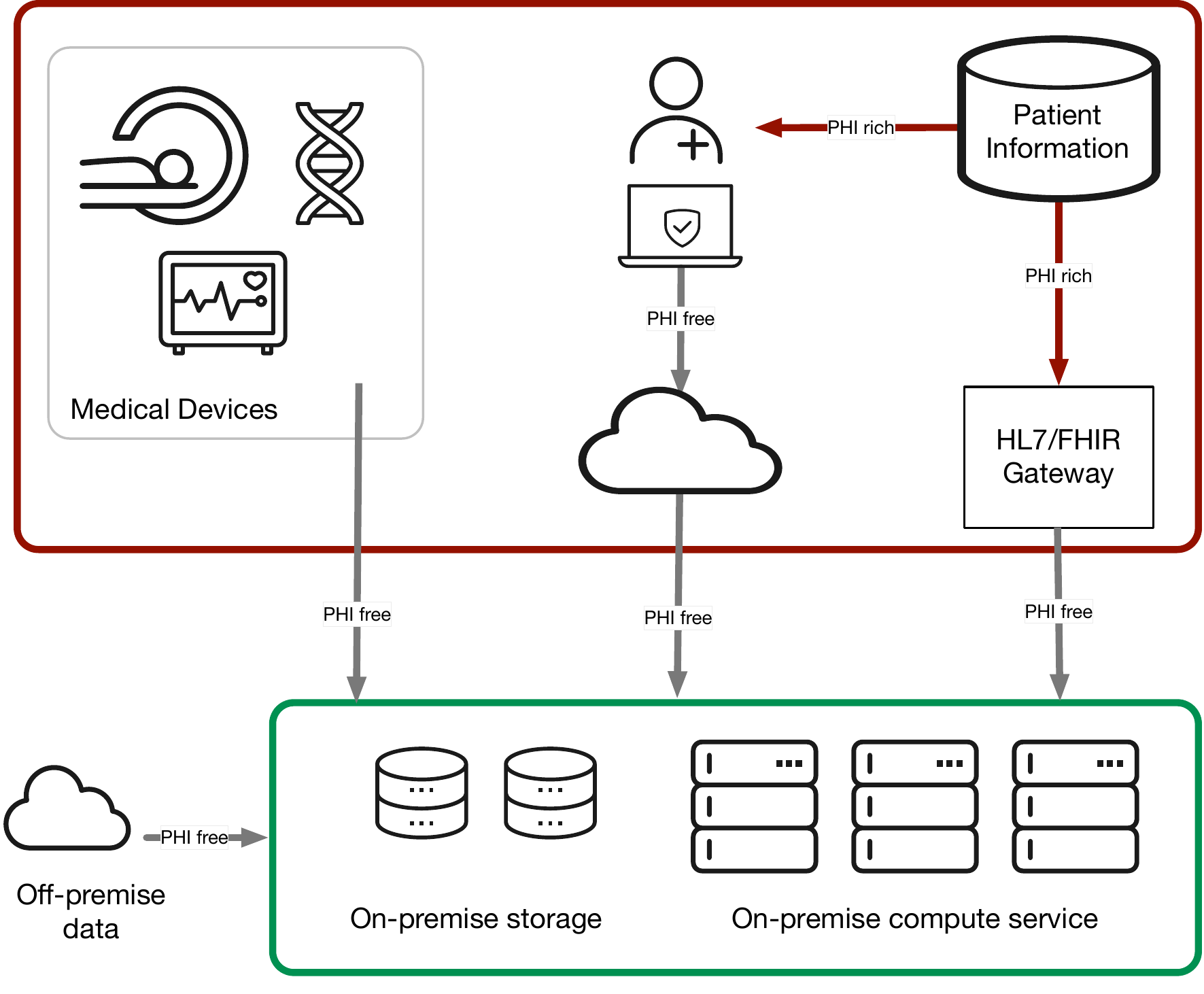}
    \caption{The data flow into the research enclave.}
    \label{fig:data_flow}
\end{figure}

Research data enters the \ac{EMCP} on several pathways, as shown in figure~\ref{fig:data_flow}. 
As indicated, it is important that solely de-identified data is present within the platform. 
To ensure this, we i) de-identify data automatically via a \ac{FHIR} gateway and in cases a fully automatic de-identification is not possible, we ii) manually perform the de-identification. 

\begin{figure}[tbp]
    \centering
    \includegraphics[width=0.9\columnwidth]{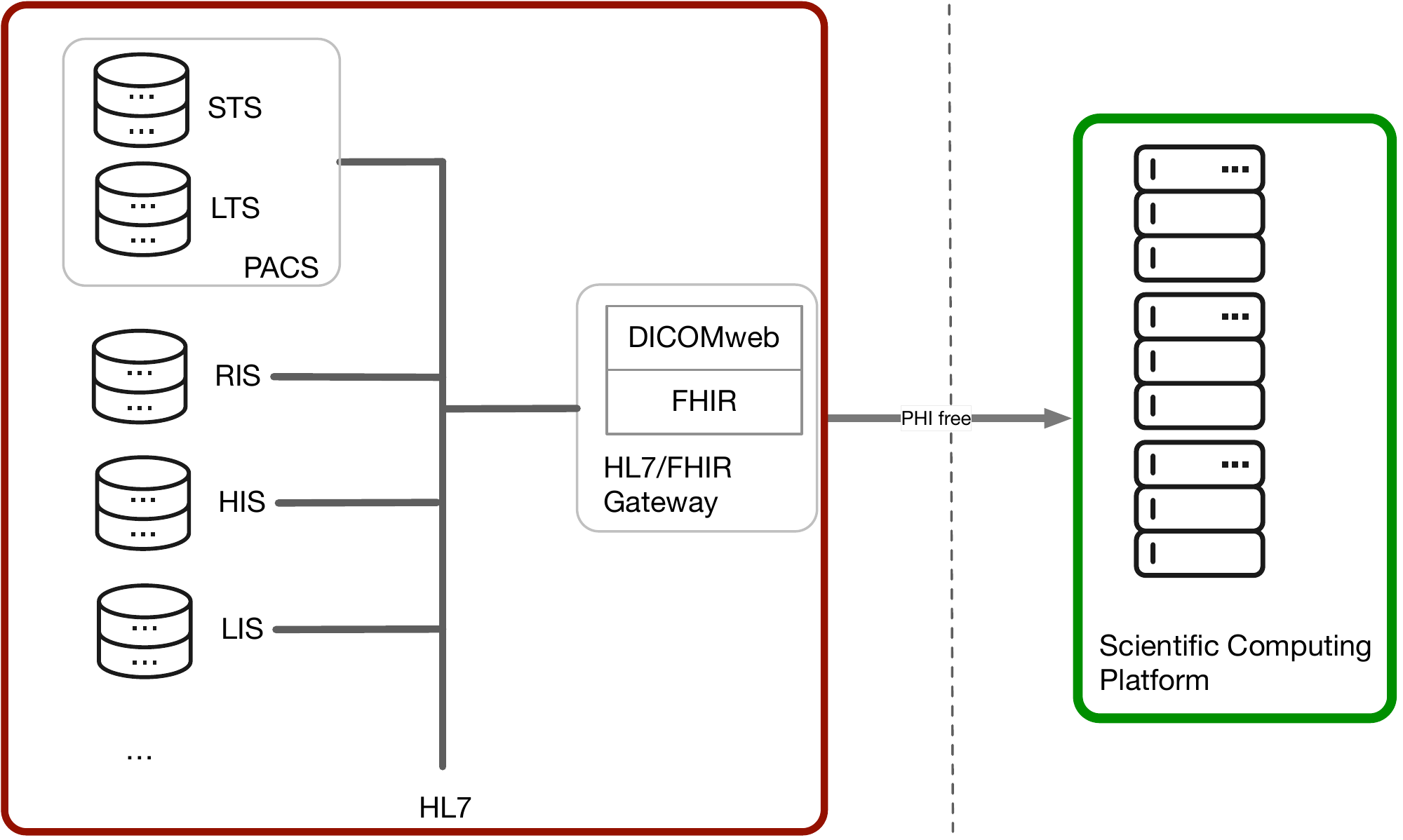}
    \caption{A dedicated \ac{FHIR} gateway is used to de-identify health care data from various data subsystems and make it available for research purposes. Abbreviations: STS - short-term storage, LTS - long-term storage, RIS - radiology information system, HIS - hospital information system, LIS - lab information system.}
    \label{fig:FHIR_gateway}
\end{figure}

\begin{itemize}
    \item \textbf{Structured health care applications:} Access to de-identified patient data happens via an authenticated pre-existing local \ac{FHIR} gateway that includes a \ac{DICOMweb} interface. \autoref{fig:FHIR_gateway} illustrates the concept.
    \item \textbf{De-identified research data:} Access to research data happens via either dedicated solutions like a dedicated research \ac{PACS} implemented with Orthanc~\cite{Jodogne2018} or enclave-hosted \ac{SMB} file services via dedicated \acp{VLAN} for laboratory devices such as DNA sequencers. We note that human genome data is a special class of data that we do not consider in this discussion.
    \item \textbf{One-offs special cases:} For one-time data dumps and other incoming transfers from the hospital ecosystem into the research enclave we provide a client for the hospital's on-premise object store.
    \item \textbf{Global incoming data:} All other data transfer cases are taken care of by the use of an S3-compatible object storage service, implemened via MinIO~\cite{minio}. The service is available inside and outside the enclave but does require strong multi-factor authentication.
\end{itemize}

\section{From a User's Perspective}
\begin{figure}[tbph]
    \centering
    \includegraphics[width=\columnwidth]{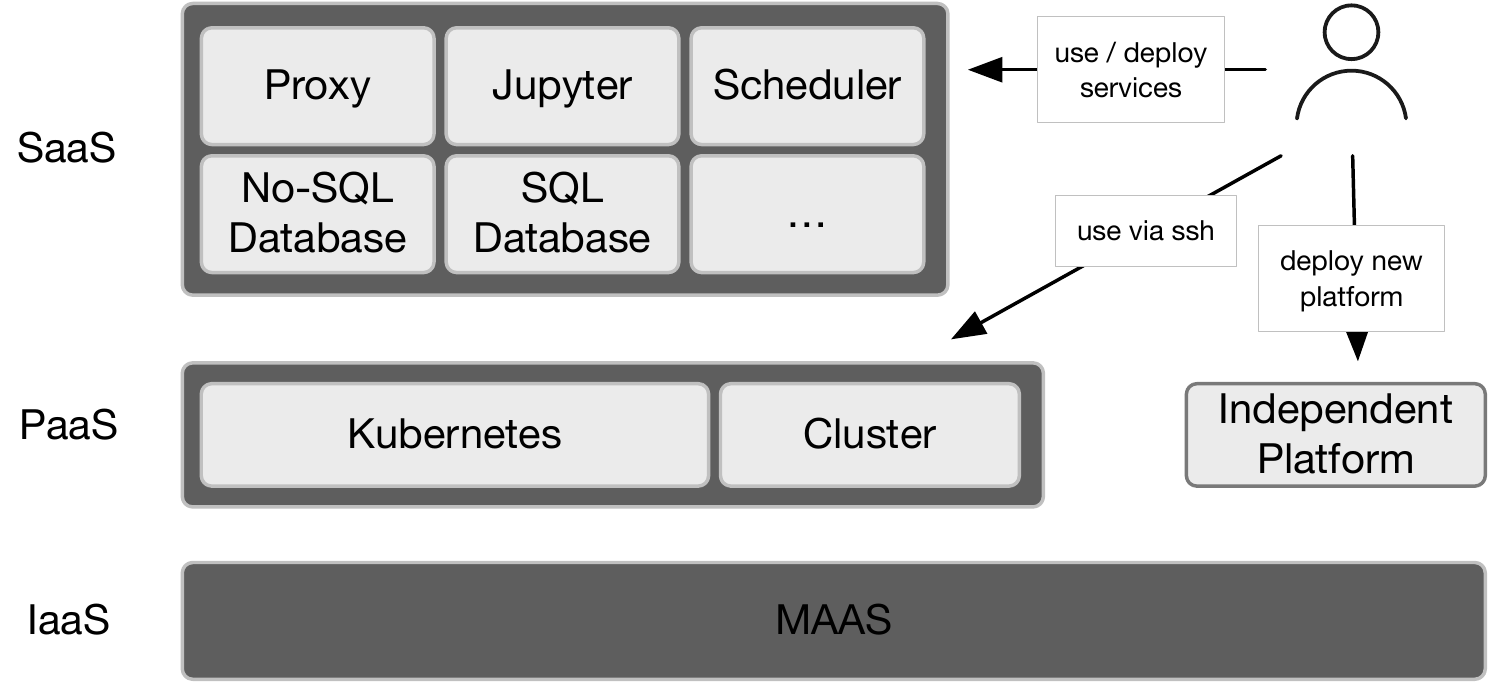}
    \caption{To the researcher the system presents as a cluster with optional services and the option to resort to the infrastructure level to deploy an indepent platform. Managed components are shown with a dark gray background.}
    \label{fig:layers}
\end{figure}

The platform consumers can choose between working with the infrastructure directly (\ac{IaaS}) or using one or more pre-installed, extensible platforms (\ac{PaaS}) and their software as a service (\ac{SaaS}). \autoref{fig:layers} illustrates this architecture from a user's perspective.

We note that while most users will choose the PaaS approach, it is critical to enable advanced users to seek alternative approaches on their own. In addition, by providing the \ac{IaC} code via a source code repository, we enable pull requests from advanced users. 
The \ac{EMCP} at its core provides access to bare metal computing resources that at first glance would appear as a traditional research cluster. The layer just above this bare metal orchestration provided by \ac{MAAS} demonstrates its power via the ability to be quickly reconfigured for various scientific workloads in a programmatic, repeatable fashion through the use of scripted infrastructure on top of \ac{MAAS}. Additional flexibility comes from rootless Linux container capabilities that allow the use of pre-defined binary environments for scientific computing, creating instantly repeatable environments to aid in the scientific process.

\begin{figure}[tbhp]
    \centering
    \includegraphics[width=0.95\columnwidth]{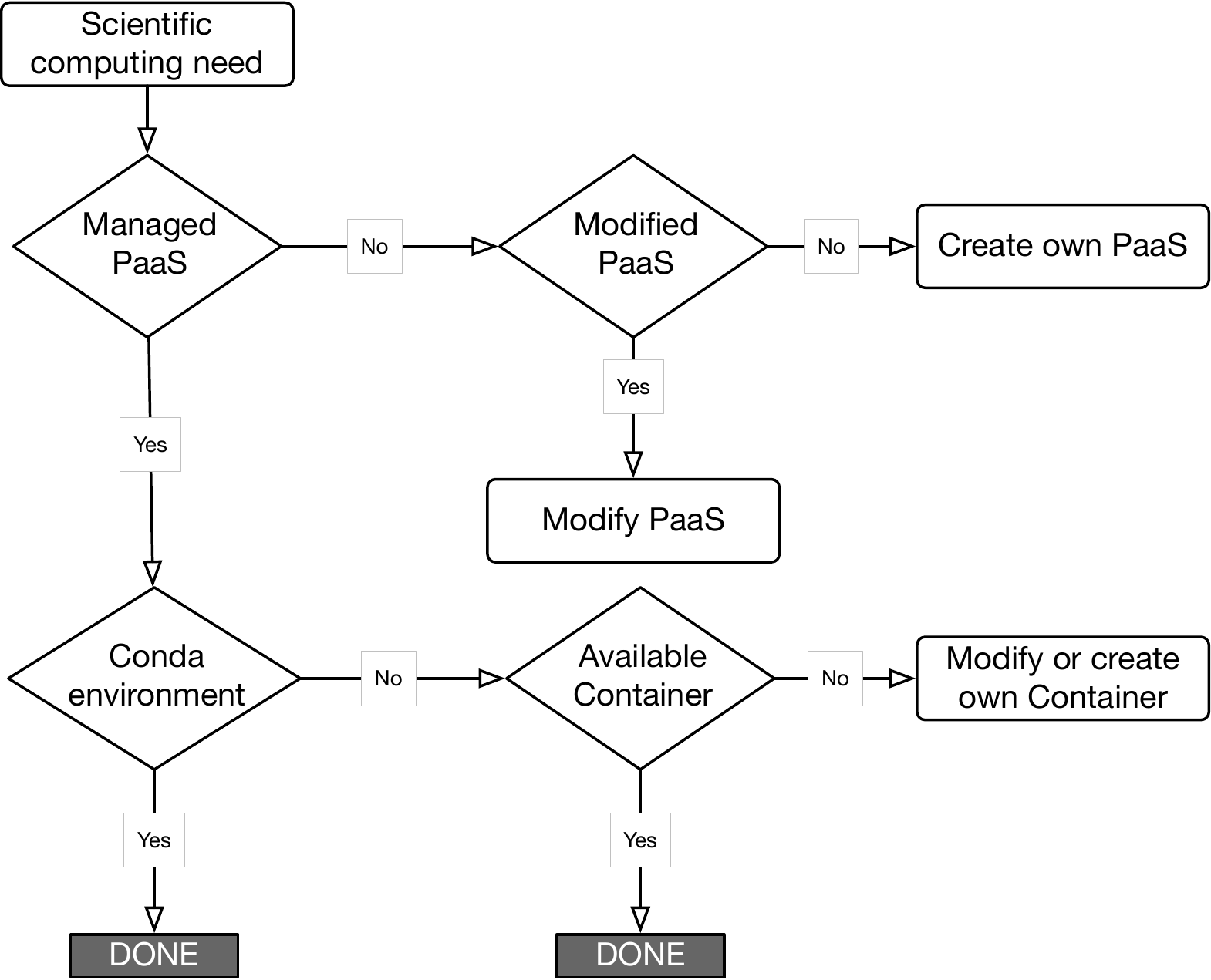}
    \caption{A wide variety of scientific computing requirements are served with the setup we provide. The graphic shows the decision process from the users' point of view. Decision nodes determine whether the scientific computing need can be fulfilled by the indicated element.}
    \label{fig:user_choice}
\end{figure}

\autoref{fig:user_choice} outlines the choices of a researcher with a scientific computing need. 
A pure-Python computing need, requiring only common libraries allows a zero-configuration quickstart via the managed \ac{PaaS} and Conda base environment.
Similarly, computing needs that can be catered by packages available via Conda can be fulfilled with minimal additional configuration as Conda environment management is available in the managed \ac{PaaS}.
We plan to simplify access even further by a JupyterHub~\cite{jupyterhub} instance, making Jupyter~\cite{jupyter} notebooks internally available without requiring command-line skills.
In particular students are familiar with such environments (e.g., Google Colab~\cite{colab}), as they rarely own \ac{ML} hardware nor have access to their institutions' facilities.

If additional software beyond a Conda environment is required, it can be deployed by using readily available containers or by setting up new ones.
If containers cannot cater the researcher's requirements, the researcher can choose the \ac{IaaS} to instantiate their own nodes, deploy a modified version of the platform or start from scratch. 

\section{Conclusion and Discussion}
We believe we have successfully established our computational research enclave and evidence suggests that research productivity is growing.

A critical component is separating research computing from patient care IT. It is this separation of concerns that enables the freedom required to conduct research, e.g., giving researchers the freedom to install and use tools without being limited by legitimate hospital IT security concerns.

Possibly the most critical point of all is that all parties trust the data privacy rule set, the operational model and agree with the implemented restrictions.

This infrastructure is a work in progress, and by design it allows newly emerging computational paradigms and/or software to be easily accommodated. More importantly, the process of change is de-centralized (or rather crowd sourced) by sharing the infrastructure as code of the platforms.

We think that lowering the barrier to access and flattening the learning curve is critical to help, e.g., students be productive with the system.
Future user modalities are likely to include readily accessible web-based workspaces via Gitpod~\cite{gitpod}.
We will continue to explore options for use of our system and implement them as needed.

\section*{List of Acronyms}
\begin{acronym}[]
 \acro{2FA}{Two-Factor Authentication}
 \acro{ASIC}{Application-Specific Integrated Circuit}
 \acro{CUDA}{Compute Unified Device Architecture}
 \acro{DL}{Deep Learning}
 \acro{DICOMweb}{Digital Imaging and Communications in Medicine Web Services}
 \acro{EHR}{Electronic Health Record}
 \acro{EMCP}{Essen Medical Computing Platform}
 \acro{FHIR}{Fast Healthcare Interoperability Resources}
 \acro{FPGA}{Field Programmable Gate Array}
 \acro{GDPR}{General Data Protection Regulation}
 \acro{GE}{Gigabit / second Ethernet}
 \acro{GPU}{Graphics Processing Unit}
 \acro{IaaS}{Infrastructure as a Service}
 \acro{IaC}{Infrastructure as Code}
 \acro{LDAP}{Lightweight Directory Access Protocol}
 \acro{MAAS}{Metal as a Service}
 \acro{ML}{Machine Learning}
 \acro{NFS}{Network File System}
 \acro{PaaS}{Platform as a Service}
 \acro{PACS}{Picture Archiving and Communication System} 
 \acro{PHI}{Protected/Personal Health Information}
 \acro{SaaS}{Software as a Service}
 \acro{SMB}{Server Message Block}
 \acro{VLAN}{Virtual Local Area Network}
\end{acronym}

\bibliographystyle{IEEEtran}
\bibliography{literature.bib}

\onecolumn

\section{Appendix: Description of Hardware in the Enclave}

The authors thought it useful to provide an overview of the component bits of the enclave as well as a list of the services established. 

Our guiding principle was KISS\footnote{Abbreviation for the design principle ``keep it simple, stupid".}. We assume that complexity will arise no matter what and by deliberately keeping the components simple we can reduce the potential for complex errors. We chose 1 \ac{GE} and 10 \ac{GE} network ports and components over potentially faster but more expensive and/or more complex alternatives, dedicated low-cost network hardware over multi-purpose, complex setups supporting multiple VLANs per device.

\begin{table}[ht]
     \caption{The enclave hardware shopping list}
    \label{tab:my_label}
    \setlength\extrarowheight{4pt}
    \centering
    \begin{tabularx}{\textwidth}{+l^p{5cm}^X}
    \toprule\tabhead
         Name & Short description & Function \\\otoprule
         Network switches & Layer 2 ethernet switches (10 \ac{GE} or faster) & Top-of-rack Layer 2 switch providing internal enclave network. The switches are cross-connected via 100 \ac{GE}. We chose traditional ethernet over fibre.\\

         Baseboard Management Controller switches & Layer 2 ethernet switch (1 \ac{GE} or slower) & Remote management out-of-band using Baseboard Management Controllers, dedicated device for security and to minimize overall complexity.\\

         Infrastructure nodes & Dual CPU Servers with approx. 200GB RAM and local redundant storage & External connectivity via proxy, reverse proxy and bastion host, \ac{IaaS}, \ac{LDAP} for auth\\ 

         File servers & Dual CPU servers with approx. 800GB RAM, 10 \ac{GE} and local storage & \ac{NFS}, SMB, S3 services \\

         Compute-Servers & Dual CPU servers with approx. 200GB RAM and dedicated local data cache respectively & CPU-bound activities, prefer many independent IO pipelines in small machines over fewer pipelines in fewer larger machines \\

         GPU Servers & Dual CPU servers with 1TB or more RAM, 6 or more \acp{GPU} and large local NVMe data cache & Machine Learning, Deep Learning\\
    \bottomrule
    \end{tabularx}

\end{table}

\begin{table}[ht]
    \caption{Current list of internal services}
    \label{tab:internal-services}
    \centering
    \begin{tabular}{+l^l^l}
    \toprule\tabhead
         Name &  Purpose                    & Comment \\ \otoprule
         \ac{MAAS} & \ac{IaaS} platform          & Ubuntu \ac{MAAS} allows user driven OS deployments on bare metal \\ 
         \ac{LDAP} & Auth and single sign-on     & implemented by FreeIPA \\ 
         Slurm & Batch processing and queueing system & \\ 
         \ac{NFS}   & industry standard shared file system & performance issues are compensated for with local disk caching\\
         \ac{SMB}   & industry standard protocols for accessing shared file systems & used for connecting to laptops and lab devices \\ 
         S3   & industry standard storage API & implemented via MinIO \\ 
         Kubernetes & industry standard for container orchestration platform & implemented via Rancher\\\bottomrule
    \end{tabular}
\end{table}

\end{document}